\documentclass{article}
\usepackage{spconf,amsmath,graphicx,hyperref}
\usepackage{lipsum}
\usepackage{cite}
\usepackage{multirow}
\usepackage[font=small]{caption}    
\usepackage{subcaption} 
\usepackage{stfloats}

\title{A Stem-Agnostic Approach to Hybrid AI Music Detection}
\name{Richa Namballa~$^{1,2}$\sthanks{This work was completed as part of an internship at Deezer Research.}, Fran\c{c}ois Rigaud~$^2$, Romain Hennequin~$^2$}
\address{$^1$~Music and Audio Research Laboratory, New York University, USA\\
$^2$~Deezer Research, Paris, France}
\begin{document}

\ninept
\maketitle
\begin{abstract}
The inclusion of generative audio in the music production process has led to an increase in hybrid music tracks that blend authentic human performances with AI-generated stems, challenging traditional AI music detectors which operate in a binary setting.
In this work, we propose a stem-agnostic framework for identifying synthetic audio sources within hybrid musical mixtures.
We introduce the inspectrogram, a novel time-frequency representation that maps localized probabilities of synthetic content across the audio spectrum.
By combining the inspectrogram with a Wiener filter estimating target stem energy dominance, a single CNN model evaluates whether the specific stem is generated.
Trained on rendered hybrid mixtures and evaluated across various stem classes, our model achieves strong performance on high-frequency sources such as vocals, drums, and guitar, but struggles on the low-frequency, narrow-band bass.
We conclude that the quality of separation impacts the detection accuracy and identify source separation as a primary bottleneck and a crucial direction for future research.

\end{abstract}
\section{Introduction}
\label{sec:intro}

The rapid spread of generative audio models has made the creation of AI-generated music more accessible than ever.
Users are producing and sharing AI content at unprecedented rates, a trend further accelerated by recent partnerships between generative music companies and major record labels~\cite{WMG2025, Hiatt2026}.
However, the popularity of these tools introduces significant challenges for commercial streaming services and creators alike, with platforms now reporting that AI-generated audio constitutes the majority of daily new music uploads~\cite{Wendel2026}.
Therefore, reliably identifying AI-generated tracks has become essential to protecting listeners and artists through actions such as clear content labeling, exclusion from recommendation algorithms and editorial playlists, and demonetization of fraudulent streams.

While existing generative music detection models perform exceptionally well on fully AI-generated tracks (often achieving accuracies above 99\%~\cite{Afchar2025icassp, Afchar2025ismir, Rahman2025}), their binary character struggles in increasingly common production settings where synthetic audio is mixed with authentic human performances.
As creators integrate generative tools into broader production workflows, the proportion of AI-sourced content within a single song varies significantly.
These ``hybrid'' human-AI tracks pose a critical challenge to current detection pipelines, as the generative artifacts these methods rely on become far less salient when blended with bonafide material~\cite{Rigaud2026, DeLaCruz2026}.
Furthermore, even when a hybrid track is correctly flagged, global detection models provide no insight into which specific components or instruments were generated versus real.

To address these limitations, we present a stem-agnostic model to identify AI-generated instrument sources within hybrid tracks.
Our framework introduces the \emph{inspectrogram}, a time-frequency representation of synthetic content probabilities, and 
processes it alongside the target stem's Wiener filter, leveraging the assumption that regions of high stem energy should strongly correlate with detection confidence in the inspectrogram.
When evaluated across various sources, including instrument classes unseen during training, model performance varies substantially by stem type.
While the system excels on high-frequency stems, it struggles to accurately detect bass due to critical spectral energy being discarded in the source separation process. 
Ultimately, the model remains constrained by source separation performance, establishing separation quality as both a critical bottleneck and a key focus for future work.

\vspace{-0.5em}
\section{Related Work}
\label{sec:related_work}

Current audio-based AI music detection systems primarily rely on identifying artifacts in a generated waveform that originate from the transposed convolution layers of a generative model's decoder architecture~\cite{Afchar2025icassp, Afchar2025ismir}.
These artifacts typically manifest as periodic spectral peaks (``checkerboard'' artifacts in the image domain~\cite{Odena2016}) that detection models learn as a discriminant signal.
While highly effective, these models are generally trained in a binary setting: assuming tracks are strictly fully generated or entirely human-made~\cite{Crosvila2025}.
Furthermore, they suffer from constrained generalization, as the models must be trained to identify artifacts specific to individual generation platforms and codecs.
  
Existing literature remains limited in addressing the hybrid music scenario where a track blends human performance with generated audio. Rigaud et al.~\cite{Rigaud2026} tackle this challenge in a two-stem setup (vocals and accompaniment) using encoded stems from MUSDB18-HQ~\cite{Rafii2019}.
They demonstrate the poor performance of a naive baseline approach which first applies a music source separation (MSS) model~\cite{Rouard2023} to a hybrid mixture and then uses a standard binary detection model on the separated stems, as MSS algorithms fail to isolate the artifacts alongside the generated stem.
To address this, they propose a multilayer perceptron model that combines local-level binary detection outputs with the relative signal-to-noise ratio (SNR) of the target stem across frequency bands, utilizing the MSS model solely to compute the SNR.
Ultimately, they conclude that detection performance depends heavily on the relative energy of the generated stem within the mix as this influences the artifacts' saliencies.

An alternative approach, proposed by de la Cruz et al.\cite{DeLaCruz2026}, adapts the binary AI-music detection task to estimate an energy ratio representing what proportion of the track is generated.
They find that the performance of the detector is stem- and frequency-dependent.
Specifically, AI-generated drums and guitar are easier to detect than vocals and bass as their codec-based artifacts have a stronger presence.
The works by \cite{Rigaud2026} and \cite{DeLaCruz2026} provide a strong foundation for our contributions in detecting AI-generated stems in hybrid mixture, as we iterate on \cite{Rigaud2026} by introducing a more fine-grained approach to detection based on both time and frequency.

\begin{figure}
    \centering
    \includegraphics[width=\linewidth]{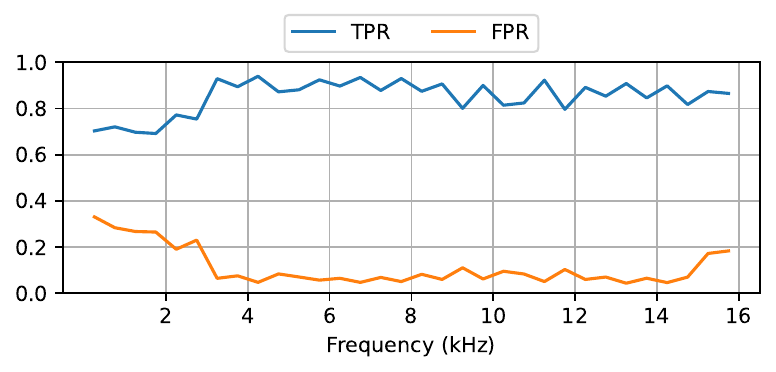}
    \vspace{-1em}
    \caption{The band-wise TPR and FPR of the BS-Reg ensemble model trained on 5-second audio segments using 500~Hz frequency bands. The reported metrics are from the test set of the FMA dataset using EnCodec for synthesizing generated mixes.}
    \vspace{-1em}
    \label{fig:bsreg}
\end{figure}

\begin{figure*}[b]
    \centering
    \includegraphics[width=\linewidth]{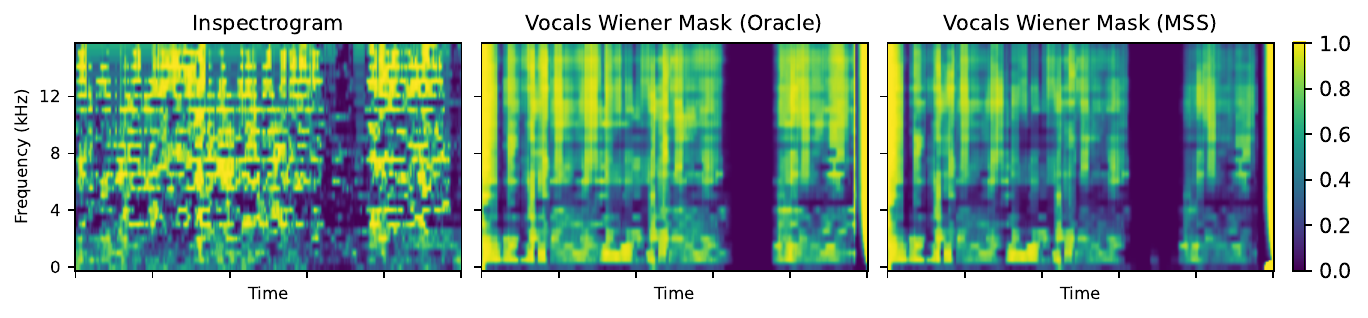}
    \vspace{-2em}
    \caption{The inspectrogram (left) of a hybrid mix  of ``Angelsaint'' by Triviul from MUSDB where the vocals are a generated stem. The energy masks of the oracle vocal stem (middle) and vocal stem separated by HTDemucs (right) demonstrate that when the vocals are active, the inspectrogram shows a high probability of synthetic content.}
    \label{fig:inspectrogram}
\end{figure*}

\section{Datasets and Features}
\label{sec:dataset}

\subsection{Hybrid Music Datasets}

While the proliferation of AI music models has led to the creation of several datasets for binary detection~\cite{Rahman2025, Zang2024, Comanducci2025, Li2026, Lopez2026icassp, Lopez2026ismir}, no public dataset currently exists for hybrid music constructed from commercial services.
Following the methodology in \cite{Rigaud2026} and \cite{DeLaCruz2026}, we render a hybrid dataset from open music source separation (MSS) corpora~\cite{Rafii2019, Pereira2024} by auto-encoding individual stems through EnCodec~\cite{Defossez2022} (24~kbps at 48~kHz).
As demonstrated in \cite{Afchar2025ismir}, EnCodec's decoder introduces architectural artifacts similar to those produced by full generative models. 
Leveraging this neural codec as a proxy for generation allows us to isolate artifact detection from confounding variables, such as semantic distribution shifts or file encoding discrepancies.
By pairing identical stems in both original and encoded states, we ensure the detection model discriminates based on synthetic artifacts rather than variations in generative content.

For training our hybrid-detection model, we construct a hybrid variant of MUSDB18-HQ~\cite{Rafii2019} using its four-stem format: vocals, drums, bass and ``other'' (VDBO).
For each of the 150 tracks, we utilize all $2^4 = 16$ possible combinations of original and synthetic stems, including the fully real and fully generated mixes.
To evaluate our model's performance on stem types not explicitly seen during training, we also curate an out-of-distribution test set using MoisesDB~\cite{Pereira2024}.
MoisesDB's 240 songs each contain varying amounts of sources, so we down-mix each multi-track into six standardized target stems: VDBO plus guitar and piano.
For each song, we randomly select $k \in [0, K]$ stems to encode (where $K$ is the number of available stems), producing a single mix per track.

\vspace{-0.75em}
\subsection{Inspectrogram}
To train our model, we introduce a novel time-frequency (T-F) data representation.
We first construct a local binary detector trained to distinguish fully authentic tracks from fully synthetic ones.
Following the setup in \cite{Afchar2025icassp, Afchar2025ismir}, we auto-encode audio from the Free Music Archive (FMA) dataset~\cite{Defferrard2017} using EnCodec to generate paired real and synthetic samples.

Rather than classifying tracks globally, we perform detection at a granular, band-limited level.
Specifically, we extract the \emph{fakeprints}~\cite{Afchar2025ismir} of 5-second audio segments (``chunks'') with a 0.25-second hop size across the 0 to 16~kHz spectrum.
We partition each fakeprint into 32 uniform 500~Hz frequency bands and train an independent logistic regression classifier per band, an ensemble we call Band-Split Regression (BS-Reg).
Each model outputs the localized probability that its corresponding frequency band contains synthetic content.
Testing various combinations of chunk duration and bandwidth showed that larger bandwidths and longer temporal segments improve accuracy, but at the expense of granularity.
Therefore, our proposed configuration provided the optimal balance between performance and T-F resolution.
Figure~\ref{fig:bsreg} demonstrates that the BS-Reg model performs reliably from 3~kHz upward.
However, its efficacy degrades at lower frequencies, with an increase in false positives and decrease in true positives, producing noisy predictions due to the dense spectral information in these regions.

By applying BS-Reg across sliding 5-second segments, we capture temporal variations in detection confidence, providing a continuous two-dimensional T-F probability map: the inspectrogram (Figure~\ref{fig:inspectrogram}).
Although trained exclusively on binary mixtures, applying the inspectrogram to hybrid tracks provides a spatially and temporally localized, but slightly noisy, representation of synthetic audio in the mix.

\vspace{-0.75em}
\subsection{Wiener Filter Masks}
As established in \cite{Rigaud2026}, synthetic stems become significantly easier to detect in frequency bands where their relative energy is dominant.
Based on this observation, we hypothesize that a model can learn the relationship between source energy and synthesis probability by correlating the mixture’s inspectrogram with a two-dimensional Wiener filter representing the source-to-mixture energy ratio (the ``energy mask'').
We compute this energy mask for each source under two conditions: an ``oracle'' scenario using ground-truth stems, and a practical ``MSS'' scenario where sources are isolated from the mixture using the 4- and 6-stem versions of HTDemucs~\cite{Rouard2023}.

To construct the energy mask, we compute the Short-Time Fourier Transform (STFT) of the target source using the same frame parameters as the fakeprint (5-second audio windows with a 0.25-second hop size).
We then aggregate the spectral bins into 32 uniform bands spanning 0 to 16~kHz yielding a T-F representation of the stem's energy activity that matches the dimensions of the inspectrogram (Figure~\ref{fig:inspectrogram}).

\vspace{-0.25em}
\section{Hybrid Detection Model}
\label{sec:hybrid_model}
We formulate stem-level hybrid detection as a binary classification task using a Convolutional Neural Network (CNN) architecture consisting of a 5-layer convolutional encoder, adaptive average pooling, and a dense classifier.
The network takes in a 2-channel matrix containing a slice of the stacked inspectrogram and energy mask. Each 64 $\times$ 32 input ``window'' consists of 64 temporal chunks (approximately 20 seconds of audio) and 32 frequency bands.
The inputs are supplied with logit-transformed values~\cite{Rigaud2026} using a 5-second hop size between windows to ensure that the model sees the entire duration of the track during training.

By characterizing target sources strictly via their Wiener masks, we effectively generalize the definition of a stem to any arbitrary sub-mix isolated from the mixture.
The stem-agnostic ground-truth label $y \in \{0, 1\}$ indicates whether the target stem represented by the energy mask is generated ($1$) or real ($0$).
This formulation assumes that the target stem itself is entirely authentic or synthetic as the behavior on stems that are themselves hybrid remains a limitation.
For inference, we average the window-level predictions $\hat{y} \in [0, 1]$ to produce a single probability score at the track level.

To prevent interference from silent or near-silent audio, we filter out unreliable samples.
Specifically, we discard any window where at least 20\% of the mixture signal consists of low-energy content.
We also filter out windows where the target stem itself is virtually silent by setting a minimum threshold for the maximum value in the window's energy mask.

\begin{figure}[b]
    \centering
    \vspace{-1em}
    \includegraphics[width=\linewidth]{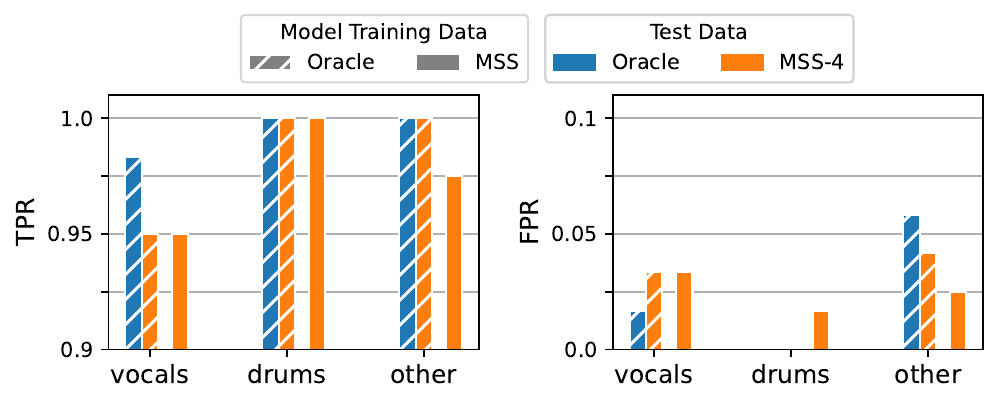}
    \caption{Track-level performance of the stem-agnostic CNN model on the MUSDB hybrid test sets using oracle stems and stem separated using HTDemucs (MSS-4). Hatched bars indicate that the model was trained on oracle stems while the solid bars specify that the model was trained on source separated stems.}
    \vspace{-1.5em}
    \label{fig:musdb}
\end{figure}

We train the CNN model using binary cross-entropy with logits loss on the rendered hybrid MUSDB dataset, using an 80/10/10 train, validation, and test split at the track level to ensure all hybrid versions of a song are in the same data partition.
Optimization runs for up to 100 epochs with an initial learning rate of $10^{-7}$ and a batch size of 256.
We employ early stopping with a patience of 10 epochs, monitoring validation performance to maximize the True Positive Rate (TPR) while also maintaining a low False Positive Rate (FPR).

Unlike prior approaches requiring a dedicated detection model for each individual stem type~\cite{Rigaud2026}, our proposed CNN is stem-agnostic: a single model can evaluate any stem in a mixture, provided the target stem's energy mask is available.
Additionally, by processing multiple chunks within a window, our model captures local temporal dependencies as well as the spectral variance leveraged in \cite{Rigaud2026}. We plan to release the code for reproducing our results.

In preliminary experiments, we trained our model on all four oracle sources in MUSDB.
While the model achieved a test set FPR under 10\% in the oracle setting, its detection performance on bass stems suffered significantly when moving to the MSS test set.
We found that HTDemucs separation heavily degrades the bass signal by filtering out higher-order harmonics and note transients.
Since bass energy is concentrated in the lowest frequency band (0 to 500~Hz), HTDemucs focuses on preserving that part of the signal, so critical discriminative information in higher bands is largely discarded during separation.
Consequently, the energy mask for separated bass becomes extremely dense in the lowest band, forcing the model to fit to the noisest bands of the inspectrogram (as seen in Figures \ref{fig:bsreg} and \ref{fig:inspectrogram}) and bias heavily toward classifying most bass stems as synthetic.

Attempting to train directly on the MSS-based energy masks did not resolve this issue, as the model still performed no better than chance on bass.
This finding aligns with the results presented in \cite{DeLaCruz2026}, where they observe that the spectral differences between generated and real stems in the bass below 5~kHz are so minor that models often struggle to discriminate between the two without access to higher-frequency information.
Based on these outcomes, we exclude bass stems from our final target stem list.
Both the real and synthetic versions of the bass stems are still included in the mixtures, but we focus the training of our AI-generated stem detector, in both oracle and MSS settings, on vocals, drums, and ``other'' (VBO) and our evaluation on those as well as the unseen guitar and piano stems.

\begin{figure}[ht]
    \centering
    \vspace{-0.75em}
    \includegraphics[width=\columnwidth]{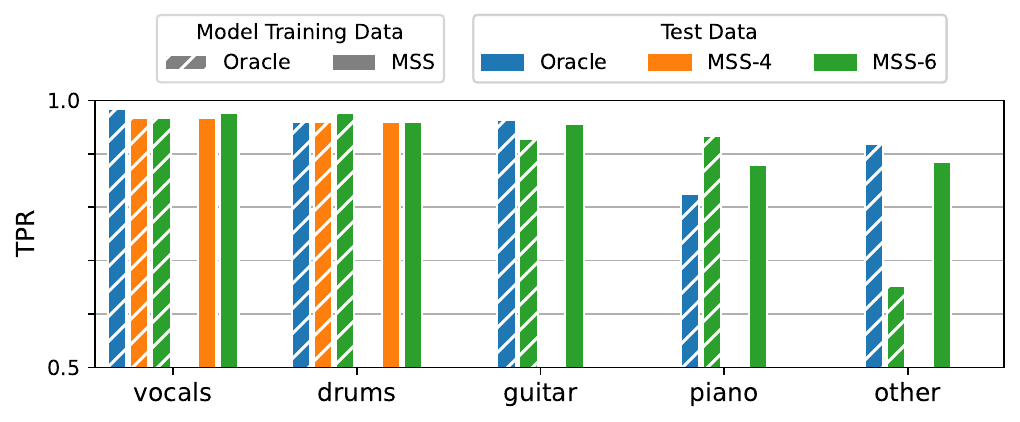}
    \vspace{-0.75em}
    \includegraphics[width=\columnwidth]{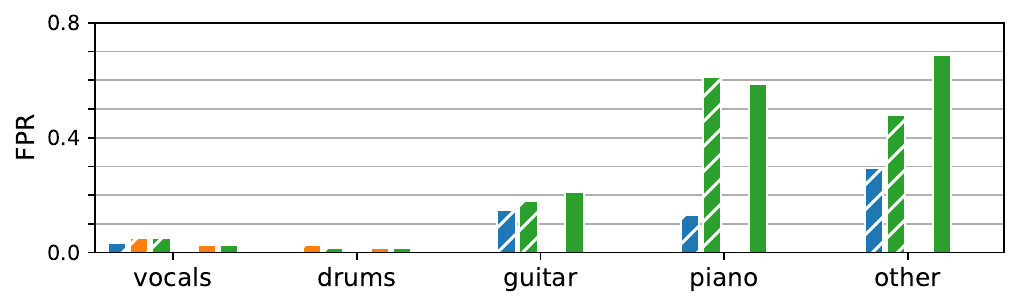}
    \vspace{-1em}
    \caption{Track-level performance of the MUSDB models trained on oracle stems (hatched bars) and on separated stems (solid bars) on the three configurations of the MoisesDB hybrid test set. MSS-4 and MSS-6 refer to the 4- and 6-stem versions of HTDemucs, respectively.}
    \vspace{-0.5em}
    \label{fig:moisesdb_results}
\end{figure}

\vspace{-1.5em}
\section{Results}
\label{sec:results}

We present the results of two CNN models: one trained on the VBO oracle stems and one trained on the VBO stems separated by HTDemucs.

When applied to the MUSDB test set (15 tracks $\times$ 16 hybrid versions), Figure~\ref{fig:musdb} demonstrates that a single model can successfully detect all three stem types from the combined inspectrogram and energy mask representations in both the oracle and MSS setting.
Interestingly, the oracle-based model performs well not only on the oracle test set, but on the MSS version as well, despite the distortion introduced by the source separation process.
One likely reason for the high performance of the oracle-based model applied to MSS-based energy masks is that HTDemucs is itself trained on MUSDB~\cite{Rouard2023}.
Lastly, moving from oracle- to MSS-based training results in only a minor performance degradation on the drums and ``other'' sources.

\begin{figure}[h]
    \centering
    \vspace{-1em}
    \includegraphics[width=\columnwidth]{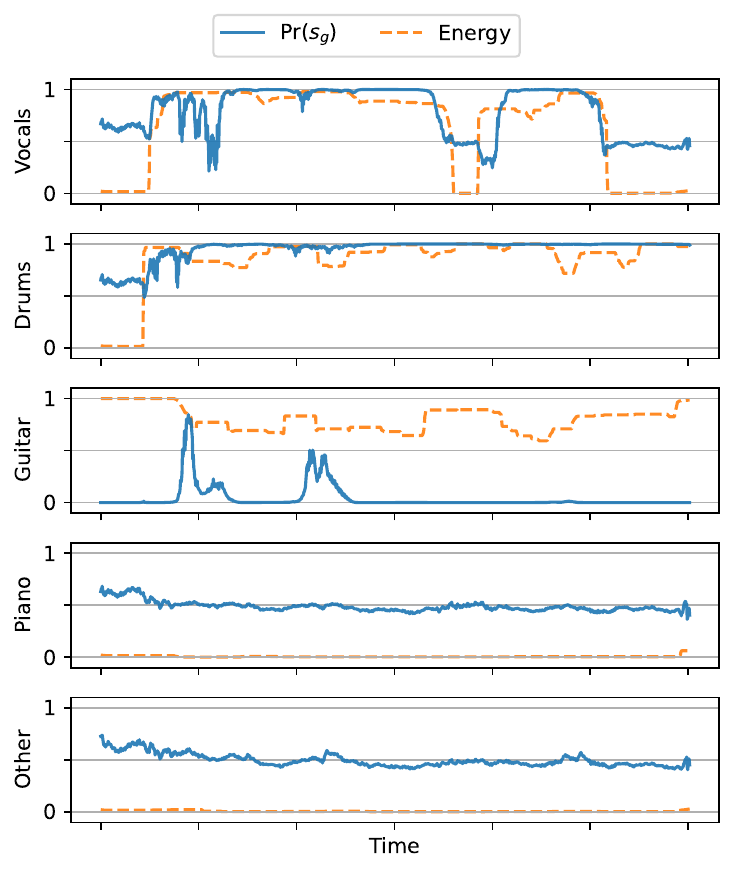}
    \vspace{-2em}
    \caption{The per-window probability that stem $s$ is generated, $\hat{y} = \Pr(s_g)$, along with the maximum value in the window's energy mask, by stem for a MoisesDB hybrid track with generated vocals, drums, and bass (not shown). The MSS-based model is applied to the MSS-6 version stems of the track.}
    \vspace{-0.5em}
    \label{fig:example_track}
\end{figure}

To better assess the generalization of our two models, we evaluate them across three test set configurations in MoisesDB: oracle ground-truth, 4-stem HTDemucs (MSS-4), and 6-stem HTDemucs (MSS-6) energy masks.
Note that we omit performance metrics for ``other'' stems under MSS-4, as guitar and piano are merged into the ``other'' category by this separator: since this sub-mix can itself be hybrid, the ground-truth label becomes ambiguous.
As illustrated in Figure~\ref{fig:moisesdb_results}, the strong detection performance on vocals and drums generalizes effectively to the MoisesDB corpus. 
Furthermore, despite not being explicitly trained on isolated guitar and piano stems (which were included in ``other'' during training), the oracle-based model identifies them relatively accurately in the oracle test set.

We notice that FPRs for piano stems increase noticeably for both the MSS-based model and in the MSS-6 test set.
Consistent with observations by \cite{DeLaCruz2026}, high-frequency artifacts in drums and guitar are inherently more salient than in piano, rendering them easier to discriminate.
Moreover, the authors of HTDemucs~\cite{Rouard2023} identify the challenge in separating piano stems and recognize that the model needs improvement on this source.
Without a reliable separated signal, our CNN model has difficulty accurately determining what parts of the inspectrogram correlate with the piano.
Lastly, in MoisesDB, piano stems are underrepresented compared to vocals, drums, and guitar~\cite{Pereira2024, DeLaCruz2026}.
When the piano does appear in the track, it is often sparse or silent across extended passages.
Low energy density in the piano source contributes to ambiguous predictions, suggesting that additional post-processing could determine whether the prediction is reliable based on the overall energy of the stem and improve overall accuracy.

Figure~\ref{fig:example_track} illustrates the direct relationship between target stem energy and the detection confidence on an example track from MoisesDB containing generated vocals, bass, and drums.
When the target stem energy is high relative to the mixture, the model yields confident predictions closer to 0 or 1 across vocals, drums, and guitar.
Conversely, during low-energy or silent passages (such as vocal pauses amid instrumental breaks), predictions hover around 0.5, reflecting uncertainty in the absence of sufficient spectral energy information to correlate with the inspectrogram.
A similar pattern is observed in the piano and ``other'' sources, where the stems are silent (or non-existent) in this track.
Without the presence of information in these energy masks, the CNN model yields ambiguous probabilities, reinforcing the need for post-processing to verify whether the stem appears in the original mixture and how much of the track it is active for.

\vspace{-1em}
\section{Conclusion}
\label{sec:conclusion}

We presented a general framework for detecting synthetic stems within hybrid music mixtures.
By introducing the inspectrogram, a T-F representation of localized generation probabilities, and combining it with target stem energy masks, our model learns to associate source energy dominance with synthetic content likelihood.
Evaluated across various sources, the system achieves peak performance on high-frequency stems like vocals, drums, and guitar.

While our experiments are based on a single neural audio codec, we believe this paradigm can be extended to work on other open-source codec architectures as well as commercial generative services, following a strategy similar to the binary detector in \cite{Afchar2025ismir}.
The amount of tuning required to adapt the system to each codec remains an open question, but we conjecture that the stem detector CNN model will work regardless of the generative model the input inspectrogram is based on.
Under this hypothesis, only the lightweight BS-Reg model would require retraining for new codecs, while the downstream CNN could be reused, provided it receives valid inspectrogram and energy mask representations.
Retraining the BS-Reg system for codec-specific inspectrograms is straightforward since collecting fully synthetic tracks from commercial models is significantly easier than assembling hybrid datasets.
Testing this adaptability provides an interesting motivation for future research.

Finally, system performance and practicality remain inherently bounded by source separation quality, as degradation in separated bass and piano stems creates a decrease in detection performance compared to the oracle setting.
Future work could explore alternative methods for computing the energy masks, such utilizing newer separation models or training lightweight, target-specific separators that directly predict low-resolution Wiener filters with the same dimensions as the inspectrogram.
Despite these limitations, our overall framework provides a scalable foundation for fine-grained, stem-level AI music detection, a increasingly critical task as co-creation between humans and AI becomes more mainstream.

\vfill\pagebreak

\bibliographystyle{IEEEbib}
\bibliography{references}

\end{document}